# Spin properties of NV centers in high-pressure, high-temperature grown diamond

O. R. Rubinas[1,2,3], V. V. Vorobyov[2,3], V. V. Soshenko[2,3], S. V. Bolshedvorskii[2,3], V. N. Sorokin[2,3], A. N. Smolyaninov[3], V. G. Vins[4], A. P. Yelisseyev[5], A. V. Akimov[2,3,6]

[1]Moscow Institute of Physics and Technology, Institutskiy pereulok 9, Dolgoprudniy, Moscow Region, Russia
[2]P. N. Lebedev Physical Institute of the Russian Academy of Sciences, Leninskiy Prospect 53, Moscow, Russia
[3]LLS Sensor Spin Technologies, Nobelev Streeet 9, Moscow, Russia
[4]LLS Velman, Novosibirsk, Russia
[5] Sobolev Institute of Geology and Mineralogy SB RAS, Prospekt Akademika Koptyuga 3, Novosibirsk, Russia
[6]Texas A&M University, 4242 TAMU, College Station, USA

## Abstract

The sensitivity of magnetic and electric field sensors based on nitrogen-vacancy (NV) center in diamond strongly depends on the available concentration of NV and their coherence properties. Achieving high coherence times simultaneously with high concentration is a challenging experimental task. Here, we demonstrate that by using a temperature gradient method of high-pressure, high-temperature growing technique, one can achieve nearly maximally possible dephasing $T_2^*$ times, limited only by carbon nuclear spins at low nitrogen concentrations or nitrogen electron spin at high nitrogen concentrations. Hahn-echo $T_2$ coherence times were also investigated and found to demonstrate reasonable values. Thus, the high-pressure, high-temperature technique is a strong contender to the popular chemical vapor deposition method in the development of high-sensitivity, diamond-based sensors.

## Introduction

Nitrogen vacancy (NV) centers in diamond are prospective solid-state systems for quantum communication and quantum computing technology [1]. Their long coherence time at room temperature makes them an interesting object for quantum information processing [2] and quantum communications applications [3]. In addition, NV centers are very attractive for metrological and biological applications. More specifically, NV centers could be used as sensitive elements for biocompatible thermometry [4] and magnetometry [5], high-resolution magnetic imaging [6] and nuclear magnetic resonance (NMR) imaging [7]. Alternatively, ensembles of NV centers could be used for implementation of ultrasensitive sensors of magnetic and electric fields [8] or strain [9].

For sensors based on NV centers in diamond, sensitivity depends on both the number of NV centers involved and on their coherence time. Unfortunately, in general, a high concentration of NV center in sample led to a short coherence time and therefore, in large enough concentrations, sensitivity often does not improve with concentration [10]. Finding the optimal balance between concentration and coherence time of NV centers requires careful optimization of diamond growing and post-processing procedures. In this work, we demonstrate that properly post-processed diamond grown by the high-pressure, high-temperature (HPHT) method, could demonstrate coherence time limited by either interaction with $^{13}C$ at a low concentration of nitrogen [11], or by interaction with nitrogen electron spins of so-called $p_1$ centers in the case of large nitrogen concentrations [12,13].

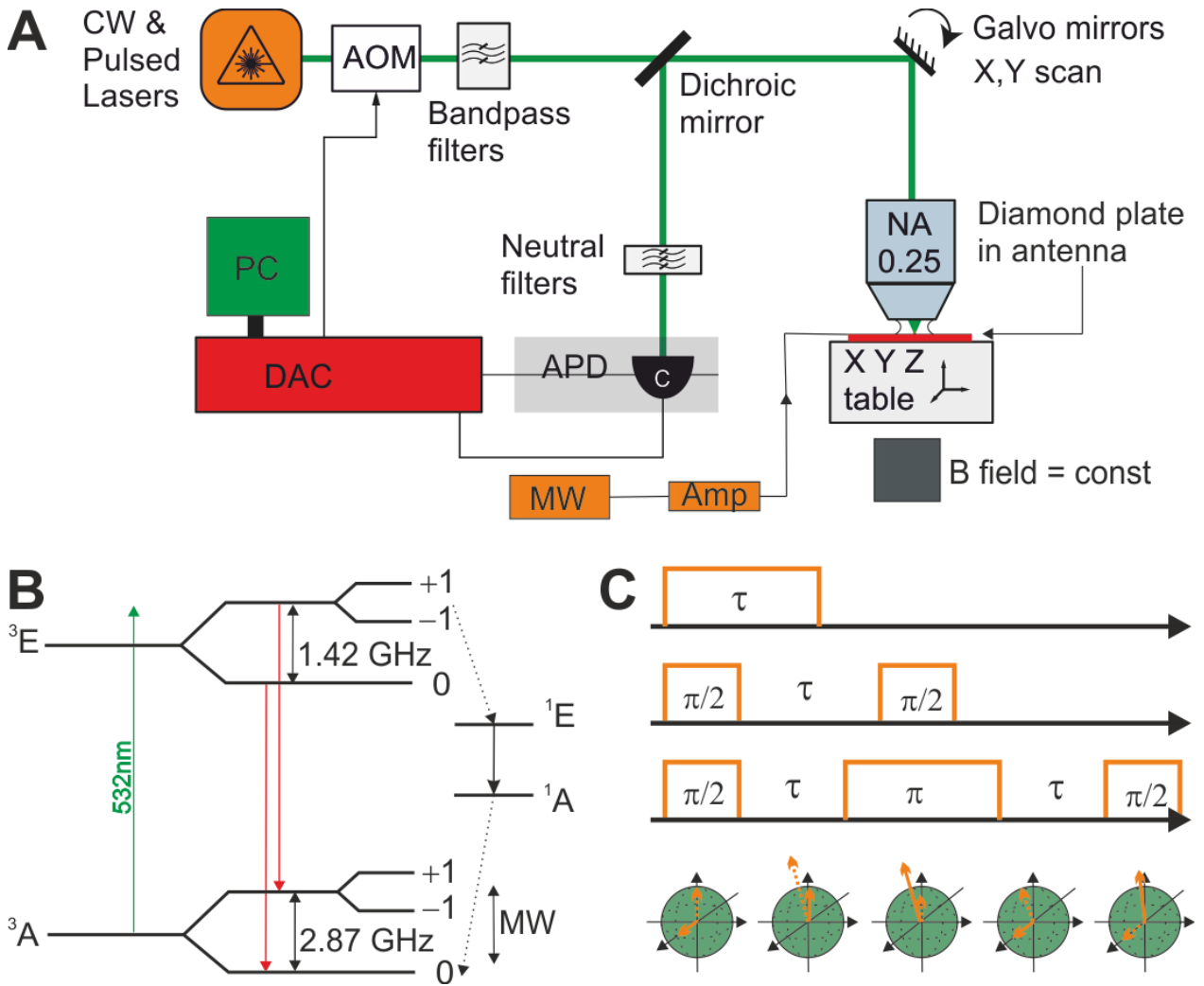

*Figure 1 A) Schematic of the experimental setup. AOM – acousto optic modulator, DAC – data acquisition card, PC – personal computer, MW – microwave source, Amp – microwave amplifier, APD – avalanche photodiode. B) Schematic of NV center energy levels. Dotted lines illustrate the non-radiative decay channel responsible for spin polarization. C) Pulse sequence used in the experiment. Top to bottom: Rabi, Ramsey, and Hahn-Echo sequences.*

## Coherence properties of diamond plates

Most of the applications of the NV color center in diamond are based on the measurement of so-called optically detected magnetic resonance (ODMR). In particular, in the case of DC magnetometry and thermometry, the microwave frequency of this resonance is the subject of measurement. Therefore, the width of this resonance, or, equivalently, its dephasing time $T_2^*$, is the crucial parameter defining the device's sensitivity.

Several factors significantly affect the coherence properties of NV centers. First, the linewidth of the ODMR resonance in ensemble of NV centers strongly depends on the strain gradient in the diamond plate and its fluctuations [14]. Correspondingly, temperature fluctuations lead to the strain gradient fluctuations or other fluctuations of phonon coupling and could affect the electron coherence time of NV center via zero field splitting $D = 2.87\,\text{GHz}$ for electron spin. In the case of nitrogen-enriched diamond, the strain is determined by the interstitial nitrogen in the crystal structure, dislocations and inclusions [15]. The next factors affecting the width of the ODMR resonance are its interaction with $^{13}$C having nuclear spin and being normally present in diamonds with a natural abundance of carbon [16], and $p_1$ defects that have electron spin on donor nitrogen [12,13]. The $^{13}$C effect represents a fundamental limitation for coherence time that cannot be overcame unless isotopically pure diamond is used [11]. The influence of defects on coherence dominates at high nitrogen concentrations and strongly depends on nitrogen-to-NV conversion efficiency. At extremely high $NV^-$ concentrations (over 50 ppm), interaction of $NV^-$ with each other may become important, as well [17,18].

Dephasing time $T_2^*$ is not uniquely defined and can be measured in different ways: by measuring the coherence decay rate in Ramsey sequence; by measuring the decay of Rabi oscillations; or by approximating the width of the ODMR resonance to zero power. Each method has advantages and disadvantages, as well as different sensitivity to the various systematics. The Ramsey sequence is quite sensitive to the hyperfine structure of NV center, causing beating in the Ramsey signal that is difficult to distinguish from decoherence. As a result, dephasing time measured by this method depends on the power of the microwaves used to excite the ground state transition [19]. At high enough microwave power, however, the power broadening caused by MW field uniformly excites all 3 hyperfine components of the ground state allowing to measure dephasing time $T_2^*$. Rabi oscillations are easier to interpret but may give a longer dephasing time than one corresponding to the width of ODMR resonance [20]. We will call dephasing time measure by this method "effective coherence time". Direct measurement of the pulsed ODMR linewidth $\Gamma_2^* \sim 1/T_2^*$ is also possible but, since linewidth is power broadened it requires set of measurement at different powers followed by fit, which find value at zero power. This method is rather time consuming therefore in our measurements we will use a Ramsey sequence coherence time as the indicator for sensor sensitivity.

In the case of AC magnetometry, sensitivity of the magnetometer could be significantly improved using the so-called Hahn-Echo technique (Figure 1C). While the Hahn-Echo technique partially removes the effect of the stray magnetic field created by diamond impurities, it is far from perfect [12]. Further improvement of coherence time could be reached with dynamical decoupling techniques that allow elongate $T_2$ to values, eventually approaching population decay time $T_1$ [21].

Nitrogen-related defects naturally accompany NV centers, but the ratio between $N_i$, $N^+$, $p_1$, etc. and NV could vary. In most cases, only a small percentage of nitrogen forms NV centers, while the rest form $N_3$ centers (three nitrogen atoms surrounding a vacancy) [22], $NV^0$ (the paramagnetic nitrogen-vacancy complex) color centers, $p_1$ (the spin containing s = 1/2 donor nitrogen atom), $N^+$ and interstitial nitrogen defects $N_i$, are usually unwanted for sensing applications due to the reasons described above [15]. Thus, a high concentration of nitrogen in the diamond leads to a high concentration of defects that affect the coherence of the NV center.

Since $T_2$ and $T_2^*$ times are ultimately limited by nitrogen concentration, it is important to maximize the yield of $NV^-$ centers. A popular way of increasing $NV^-$ yield is irradiation of diamond with electrons and subsequent annealing [23,24]. The first step creates additional vacancies necessary for the formation of $NV^-$ centers, but also generates a lot of interstitial carbon and nitrogen atoms, which could be source of strain. These interstitial carbon atoms mostly vanish after annealing at $450°C$ temperature [25,26] and vacancies not captured by nitrogen also disappear at $600°C$ [27]. At temperatures above $1200°C$, the defects of interstitial nitrogen atoms become mobile, thus additionally reducing strain. $N^+$ color centers disappear at temperatures around $1400°C$ [15]. Further increases in temperature help remove other defects in diamond, but at $1600°C$ in vacuum, diamond converts to graphite [15], which sets an upper limit on possible annealing temperatures. In principle, even higher annealing temperatures are possible if high-pressure annealing is implemented [15]. The duration of annealing is typically 2-3 hours and longer times do not provide a higher yield of $NV^-$, while they might help further reduce strain and remove unwanted defects [28].

## Results

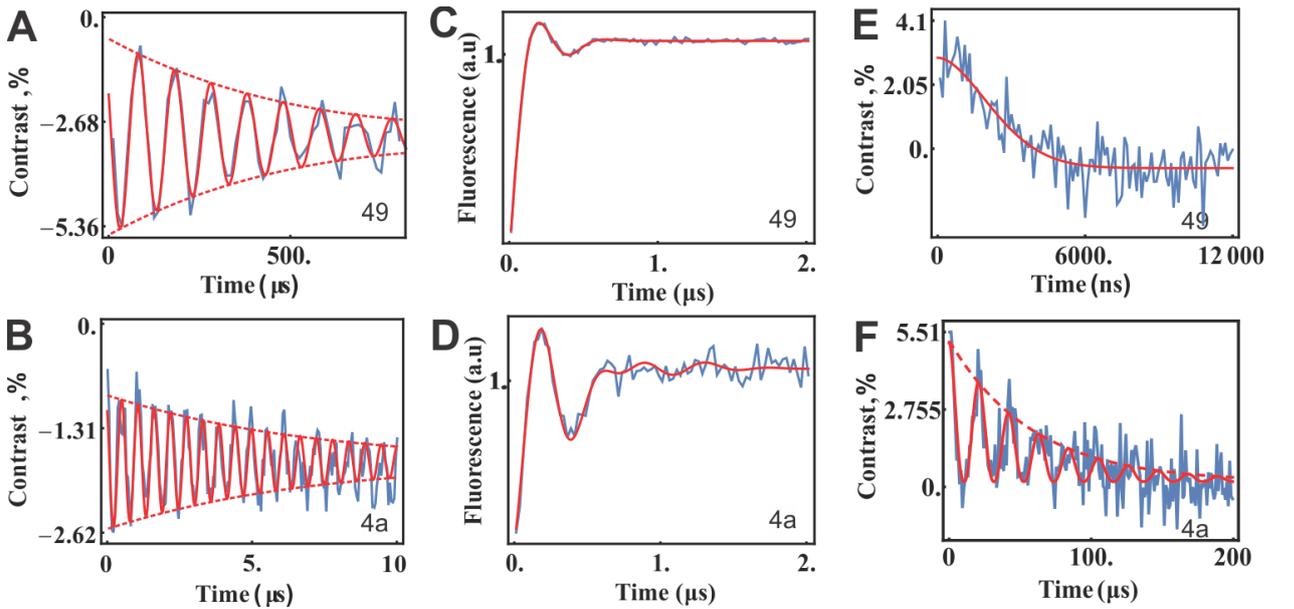

*Figure 2 A,B) Examples of Rabi oscillation for various diamond plates. Blue line stands for experimental data; solid red line for fit; red dashed line illustrates fit of $T^*_{2Raby}$ decay time.*

*C,D) Examples of measurement of $T^*_{2Ramsey}$ decay time by Ramsey sequence. Blue line stands for experimental data; solid red line for fit. E,F) Examples of $T_2$ coherence time for various diamond plates. The red solid line stands for fit. Red dashed line corresponds to envelope. The index in the bottom right corner is the plate number; see Table 1 for details.*

For measurement of the coherence properties of the diamond plates grown, we used a home-built confocal microscope [29] (Figure 1A). Control of the spin state was carried out using a microwave antenna [30] and a permanent magnetic field was applied to split the ODMR resonance to address well-defined microwave transitions between electronic spin levels $m_s = 0$ and . $m_s = 1$ . in the ground $^3A$ level (see Figure 1B and Methods). Coherence time was measured by fitting echo signal using the Hahn echo sequence (Figure 1C); the effective coherence time was obtained using the decay time of optical Rabi oscillations between the levels $m_s = 1$ and $m_s = 0$ (Figure 2A), and dephasing time was measured by Ramsey sequence (see Methods for the details) at high microwave powers (Figure 2C).

We started our studies with a relatively high concentration of nitrogen introduced on growing stage 100 ppm. Two plates with NVs concentrations were studied: 11.6 and 14.6 ppm. In both cases, measured coherence time $T_2$ was moderate and equal to 3.7 and 3.2 µs, respectively (see Figure 2E and Methods). Nevertheless, due to a very high concentration of NV centers, the potential sensitivity of magnetometer utilizing these plates could be quite high (see below). The number of NV centers in plates with high content was estimated using the low temperature absorption spectrum (see Methods). Additionally, the plate with a slightly lower nitrogen concentration had a higher decoherence times for both $T_2^*$ and $T_2$ (see Table 1). This observation is consistent with earlier works attributing $T_2^*$ to the influence of the electron spin of nitrogen (for example, one NV⁻ or $p_1$ centers) [12,13].

A second group of diamond plates with a lower initial nitrogen content (below 10 ppm) was produced. All samples were irradiated with electrons and annealed at a temperature of 800 °C or 1400 °C (see Table 1). All produced samples were measured the same way as the first run (see Figure 2B,D,F). The method of measuring ZPL in absorption spectrum is not suitable for determining NV⁻ at concentrations below 10 ppm; therefore, all plates in the second group were normalized to the plates in the first group using the slope of the saturation curves (see Figure 3A, Methods).

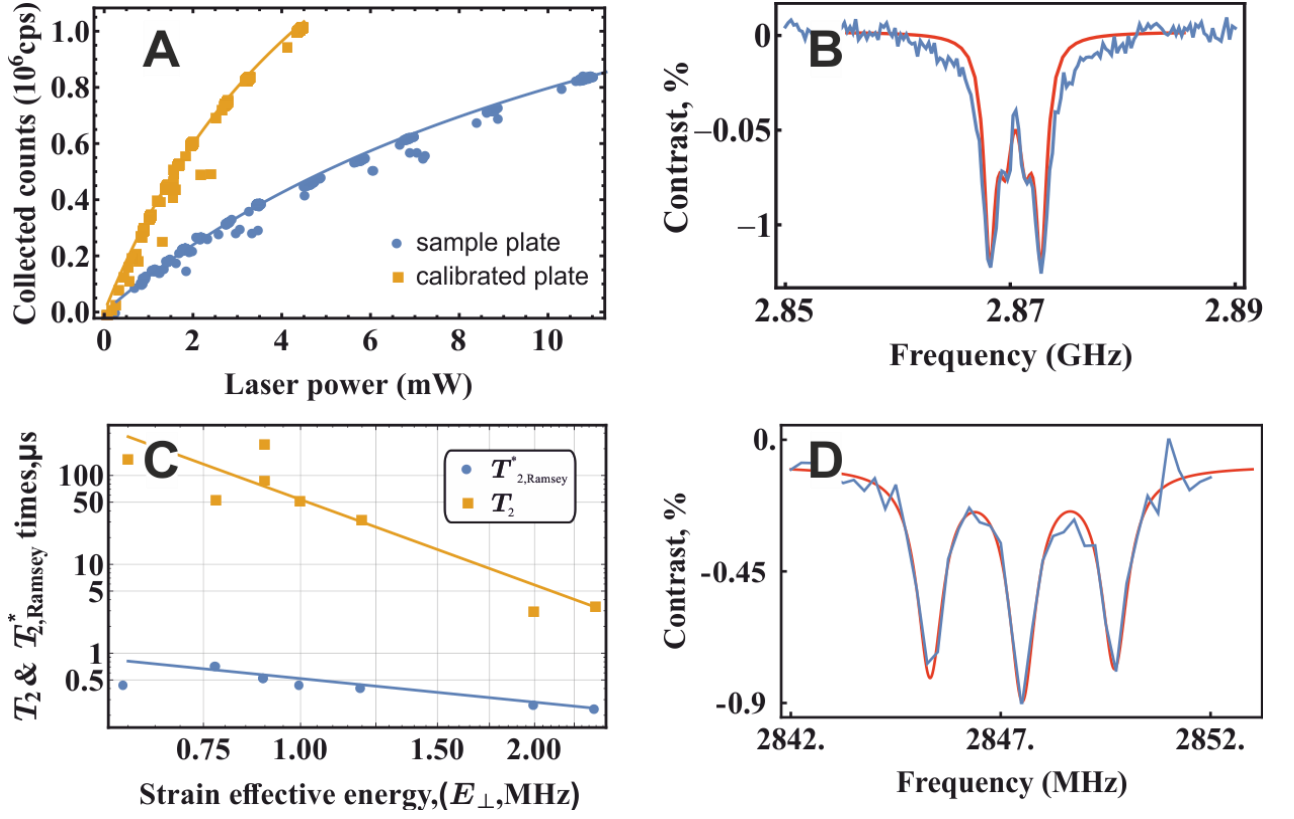

*Figure 3 A) Saturation curves for concentration measurement. Solid curve $f(x)=ax/(x+b)$ where a, b are fitting parameters. B) Splitting of the ODMR signal due to strain in the diamond plate. The red line represents the fit. C) Dependence of $T_2^*$ on strain measured in the diamond plates. The solid line represents linear fit. D) ODMR spectra for sample 4a with clearly resolved hyperfine structure. The red line represents the fit with 3 Lorentzian curves.*

While annealing considerably relieves the strain in the samples, it does not remove it completely. The strain (perpendicular to the NV axis strain component) could be estimated using the splitting of the spectral lines at zero magnetic field [26,31] (see Figure 3B). The fit of the line splitting was performed using a model of strain and magnetic field dependent ODMR spectra, obtained from [31]. It was found that the strain component strongly correlates with dephasing $T_2^*$ time, demonstrating dependence (see Figure 3C):

$$T_{2,Ramsey}^* \sim 1/E_\perp^{0.89}. \tag{1}$$

This is not surprising since strain is known to affect broadening of the ODMR line ($T_2^*$) [14].

Besides the strain component strongly correlates with coherence time $T_2$, demonstrating dependence:

$$T_2 \sim 1/E_\perp^3. \tag{2}$$

We found, that there is no direct correlation of the NV concentration with $T_2$. This could be understood as following: the $T_2$ time depends strongly on all paramagnetic impurities in the sample. For example, different regimes of electron irradiation lead to various concentration of paramagnetic impurities. Hence with less electron doses of $6 \cdot 10^{17} cm^{-2}$ for sample 4ab resulting in almost twice better $T_2$ time comparing to 2a plate, despite their NV concentration almost equal. At the same time, for plates 3a and sd2, with equal concentration of NVs and equal dose of irradiation, the $T_2$ time are similar.

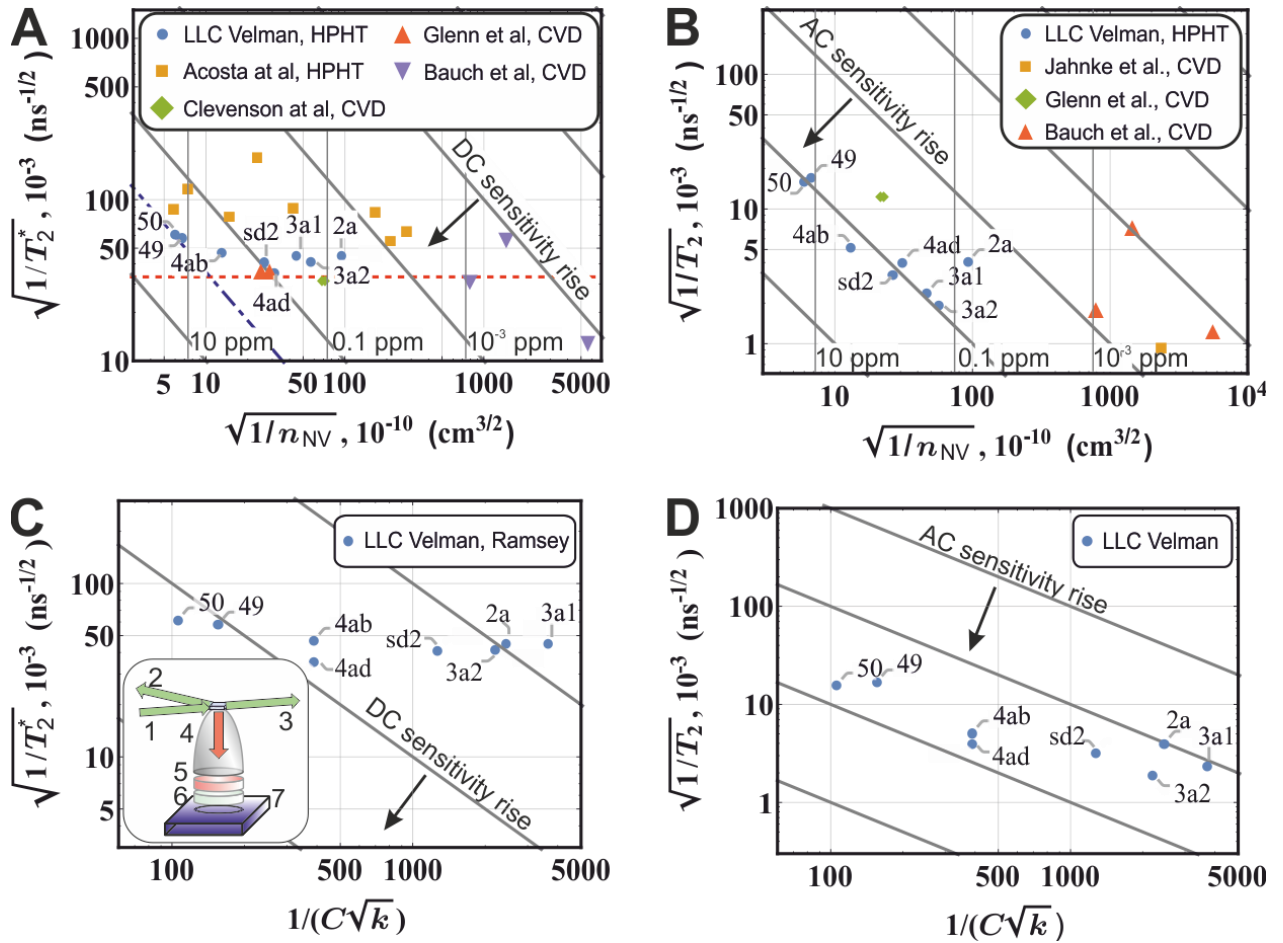

Figure 4 Coherence time $T^*_{2\,Ramsey}$ (A) and $T_2$ (B) versus NV concentration for various plates, including taken from [11,17,32–34]. Diagonal grey lines correspond to lines $\sqrt{T_2 n_{NV}} = const$. Lines are spaced with 10 times improvement in the $\sqrt{T_2 n_{NV}}$. The blue dot-dash line stands for $p_1$ limit; the red dashed line for $^{13}C$ limit. C – efficiency of the diamond plates in terms of DC sensitivity in case of fixed excitation power. Maximum sensitivity is in the origin of the coordinate system. Inset: coefficient k measurement scheme. 1 – incoming pump beam, 2 – reflected pump beam, 3 – transmitted pump beam, 4 – parabolic concentrator, 5 – long-pass filter with cutoff at 650 nm, 6 – notch filter for pump (532 nm) beam, 7 – photodetector. D – Efficiency of the diamond plates in case of fixed excitation power in terms of AC magnetometry

The best value of coherence time $T_2 \simeq 250$ μs was achieved on the sample with a nitrogen concentration of about 1 ppm; the plate was irradiated with dose $2 \cdot 10^{18} cm^{-2}$ and annealed at 1600 °C (sample 3a2). The best dephasing time was achieved on different plate, 4ad, which therefore demonstrated clearly resolved, hyperfine components (Figure 3D). In order to achieve the smallest possible width for Figure 3D, a pulsed ODMR sequence was used. To minimize the width $\pi$ pulse duration was kept large, larger than dephasing time, this way minimizing power broadening. Smaller power nevertheless leads to some decrease of the contrast achieved (approximately $1/e$ for pulse length used). Furthermore, the full contrast is divided between 3 possible hyperfine states, and thus, only contrast of $C \approx .009$ was achieved.

*Table 1 Comparison of coherence times and concentration of NV centers for various samples studied.*

|  | **49** | **50** | **2a** | **3a1** | **3a2** | **Sd2** | **4ad** | **4ab** |
|---|---|---|---|---|---|---|---|---|
| $n_{NV}$, ppm | 11.6±1.7 | 14.6±2.2 | 0.06±0.03 | 0.24±0.02 | 0.16±0.02 | 0.75±0.2 | 0.54±0.02 | 3.07±0.04 |
| $v_{irradiation}$ $10^{18}\text{cm}^{-2}$, | 1.5 | 2 | 2 | 2 | 2 | 2 | 0.6 | 0.6 |
| $t^{\circ}_{annealing}$, °C | 800 | 800 | 800 | 1400 | 1600 | 1400 | 800 | 800 |
| $T_2$, μs | 3.2 ± 0.6 | 3.7 ± 0.3 | 57 ± 4 | 167 ± 20 | 249 ± 35 | 97 ± 7 | 58 ± 8 | 35 ± 2 |
| $T^*_{2Rabi}$, μs | 0.88 ± 0.14 | 0.80 ± 0.08 | 8.04 ± 0.99 | 4.80 ± 0.65 | 3.91 ± 0.8 | 5.50 ± 1.60 | 7.02 ± 0.95 | 2.24 ± 0.40 |
| $T^*_{2Ramsey}$, μs | 0.28 ± 0.05 | 0.25 ± 0.05 | 0.46 ± 0.05 | 0.46 ± 0.06 | 0.55 ± 0.09 | 0.56 ± 0.06 | 0.75 ± 0.07 | 0.43 ± 0.05 |
| $E_{\perp}$, MHz | 2 ± 0.2 | 2.4 ±0.15 | 1 ± .1 | 0.6 ± 0.1 | 0.9 ± 0.1 | 0.9 ± 0.04 | 0.78 ± 0.06 | 1.2 ± 0.04 |
| $k$, % | 0.4 ± 0.05 | 0.6 ±0.06 | 0.01 ± 0.001 | 0.047 ± 0.002 | 0.065 ± 0.002 | 0.15 ± 0.02 | 0.18 ± 0.01 | 0.18 ±0.01 |
| $n_N$, ppm | 92±23 | 93±23 | <10 | <10 | <10 | <10 | <10 | <10 |
| $C$, % | 10 | 12 | 4 | 1 | 1 | 2 | 8 | 8 |
| $I_{NV^-}$, , % | >95 | >95 | 40 | 40 | 50 | 70 | 50 | 50 |

The results of all measurements are summarized in Table 1. Each sample was examined in magnetic field $B = 86.5 \pm 0.2$ G which was estimated from frequency of ODMR resonance using following simple formula: $|f - 2870 \text{ MHz}|/(2.8 \text{MHz}/\text{G})$, where $f$ is frequency of the ODMR resonance. The fraction of NV$^-$ fluorescence was estimated by fitting overall emission spectrum in the range 580-700 nm with combination of NV$^-$ and NV$^0$ spectra.

To better understand the potential of HPHT plates in terms of their application in a magnetometer, we also summarized our data in Figure 4. Here, key parameters such as nitrogen concentration and coherence time are plotted. The sensitivity $\eta$ to DC magnetic field to the first approximation [5] for single NV center is:

$$\eta = \frac{\hbar}{g\mu_B C \sqrt{T_2^*}} \tag{3}$$

Which for ensemble of NV center will lead to sensitivity per squire root volume $\eta_V^{DC}$:

$$\eta_V^{DC} \sim \frac{1}{C\sqrt{n_{NV} T_2^*}}, \tag{4}$$

Here $C$ is the contrast in spin dependent optical signal, $T_2^*$ – dephasing time, $n_{NV}$ – concentration of active NV centers, $g$ – NV center electronic spin gyromagnetic factor, $\mu_B$ – Bohr magneton and $\hbar$ – reduced Plank constant. Thus, the only three parameters affecting sensitivity are contrast $C$, coherence time $T_2^*$ and $n_{NV}$.

Accordingly, for AC field [5] :

$$\eta = \frac{\hbar}{g\mu_B C \sqrt{T_2}} ,. \qquad (5)$$

and

$$\eta_V^{AC} \sim \frac{1}{C\sqrt{n_{NV} T_2}} . \qquad (6)$$

To understand potential of the plates measured for magnetometry, we compared our samples in terms of coherence time and concentration of NV centers, using $1/\sqrt{n_{NV}}$ , $1/\sqrt{T_2^*}$ as the graph axes for Figure 4A, and $1/\sqrt{n_{NV}}$ , $1/\sqrt{T_2}$ for Figure 4B. Thus, Figure 4 helps to see the potential of different diamond plates under study. The way it is plotted maximum sensitivity (minimal possible to measure field) is at the origin of the coordinate system. It is clear from the picture, that both plates 49 and 50 have the best combined concentration and coherence properties. Though coherence time for this plates is moderate, ($3.2\pm0.6$ $\mu$s and $3.7\pm0.3$ $\mu$s correspondingly, see Figure 2) high concentration of NV centers in this samples ($11.6\pm1.7$ ppm and $14.6\pm2.2$ ppm, see Table 1) made this samples quite competitive not only among samples under study, but also other samples from the literature, also indicated on the graph.

To understand the limiting factors affecting coherence properties on the NV ensembles, we analyzed behavior of the $\gamma = 1/\sqrt{T_2^* n_{NV}}$ parameter from sample to sample. As can be seen in Figure 4A, in the DC magnetometer case $\gamma$ of the magnetometer already saturates at around 1 ppm of NV$^-$ concentration. Further increase of the concentration proportionally decreases coherence time and is likely due to the increase of $p_1$ centers' concentration. This limit in principle could be further reduced by increasing conversion efficiency of nitrogen to NV$^-$. At low nitrogen concentrations, coherence time seems to approach the limit caused by the presence of $^{13}$C observed by the other groups [11]. Thus, flattening of the Figure 4A at low concentration is quite expected and only could be overcome using isotopically pure diamonds.

In the case of AC magnetometer, similar behavior could be observed (Figure 4B). High nitrogen concentrations in the case of CVD plates was shown in a paper [11] to have limit for coherence time of $T_2 = 165$ $\mu$s·ppm$/n_N$, where $n_N$ is the concentration of $p_1$ centers. For the plates we examined, the maximum concentration of $p_1$ centers measured by the infrared spectroscopy method reached $68\pm17$ ppm. For these plates, the coherence time measured was $3.2\pm0.6$ $\mu$s and 3.7±0.3 $\mu$s which is, slightly better than the 2.4±.6$\mu$s limit [11]. While there is slight difference, within $2\sigma$ these results are consistent. A similar difference was observed in other work [35]. Another important point is that plates 4ab and sd2 win in sensitivity compared to 49 and 50. This is nevertheless only true if sensitivity per unit volume is the figure of merit.

In practical terms, sensitivity of the magnetometer is not limited much by the concentration of color centers, but rather by the fluorescence these color centers produce. Unfortunately, NV$^-$ color centers are not the only absorbers in the diamond; therefore, the number of emitted photons per pump photon $k$ is far from unity, even with 100% collection. To characterize the coefficient $k$, we built a test setup in which both reflected and transmitted pump beam intensity were measured, thus enabling an estimation of the absorbed pump power. The diamond was glued to a parabolic concentrator output connected to the filters and a photodetector measuring the power of the emission of NV centers (see Figure 4C inset). The efficiency of such an assembly was estimated at approximately 6% (see Methods). The results are summarized in Table 1. Surprisingly, plates with larger NV concentrations also have a larger $k$ coefficient, evidence that at low NV concentrations absorption in the samples was not limited by the NV centers. Even at high concentrations, with overall collection of only 0.6% (about 9.7% assuming perfect collection), this coefficient cannot be explained by the assumption that NV centers are the only absorbers, since NV quantum efficiency was measured at over 0.7 [36]. CVD plates demonstrate quite similar results [32]. This fact is not fully understood and requires further investigation. Additionally, the measured ODMR contrast in plates with higher concentrations tend to be also larger (see Table 1). For plates 49 and 50 the contrast reaches nearly maximum possible value for single-NV orientation [37]. While for the plate 4ab/4ad contrast is also high, other low nitrogen concentration plates demonstrate quite low contrast which even further reduces possible sensitivity of the magnetometer.

Currently the most sensitive volume-based magnetometers consume few Watts of optical power [32,38] which is already quite challenging to handle in the rather small diamond plate; therefore, it is reasonable to compare the possible sensitivity of magnetometers per root power. Assuming that optical power could be distributed over the sample in such a way that it would be fully absorbed [32], one could obtain from (4) the following dependence for total magnetic sensitivity on incident optical power $P_{laser}$:

$$\eta_{Full}^{DC} \sim \frac{1}{VC\sqrt{\alpha N_{NV} T_2^*}} = \frac{1}{C\sqrt{k P_{laser} T_2^*}} \tag{7}$$

where $\alpha$ – number of detected photons per single NV center per single measurement, $N_{NV}$ – number of active NV centers, and $V$ is the volume of the sample under laser radiation. Here we neglect service time to not limit plate performance by the specific setup and assume measurement time to be equal to the dephasing time. Similar formula could be derived for $\eta_{Full}^{AC}$. Following simple estimation (7) we will focus on only three main parameters affecting sensitivity per root power are contrast $C$, coherence time $T_2^*$ and coefficient $k_{power}$.

From Figure 4C,D one could see that the most suitable plate for a magnetometer seems to be sample №50. While in case of DC magnetometry low concentration plate on average lose in sensitivity per given power to the high-density ones, for AC magnetometry plate 4ad is quite competitive with high density ones, but have addition advantage of having narrow resolved hyperfine components. This conclusion indicates the very high potential of high-pressure high temperature growing method from point of view of diamond magnetometry making it quite competitive with commonly used CVD method.

# Conclusion

We demonstrated that diamond plates grown by the HPHT method have dephasing time $T_2^*$ and limited by unavoidable impurities such as $^{13}$C or $p_1$ centers.

The low strain of plates, remarkably high contrast to the ODMR signal were accompanied by coherence properties, naturally limited by presented $p_1$ paramagnetic impurities at high nitrogen concentration limit (sample # 49) and $^{13}$C nuclear spin impurities for low nitrogen concentration (sample # 4a dark), making HPHT diamond plates attractive candidates for the realization of ensemble-based sensors. Depending on application, either a long coherence time and spectral resolution of nuclear sidebands of ODMR signal or the best product of coherence time and concentration provided reasonable $T_2^*$ may be desired. For example, recently suggested rotation measurement using nitrogen nuclear spin in NV center may strongly benefit from resolved hyperfine component. Indeed, spectral resolution of hyperfine components will obviously allow determining the nuclear spin states in NV center ensembles thus making measurement based on nuclear spin state efficient.

The HPHT method offers both options either with long limited by interaction with $^{13}$C $T_2$ coherence time at low concentrations or $p_1$ limited high-concentration plates for high-sensitivity applications.

# Methods

## *Diamond crystal growth*

To grow and post process our samples, we used the following recipe. An HPHT multi-ton press was used, capable of producing pressure of about 5 GPa at a temperature of 1500°C. Diamond crystals were grown using the temperature gradient method in a Fe-Co-C system. The temperature gradient method makes it possible to create conditions for the stable growth of diamond single crystals on a specially introduced seed. A temperature gradient is intentionally created in the reaction zone of the high-pressure cell. The source of carbon (diamond or graphite) is placed in the region with a higher temperature, and the seed crystals of the diamond, with a lower temperature. A carbon source at a higher temperature dissolves in the melted metal, and carbon diffuses through the melted metal and crystallizes on the seed crystals of the diamond at a lower temperature [15].

In this method, the minimum rates of growth (about $10^{-5} - 10^{-6}$ cm/s) were realized. As a result, the crystals had a perfect crystal structure with a minimum content of macro- and microdefects and contained mostly nitrogen-containing impurities. Such a diamond of I b type grown by this method can contain from 1 to 100 ppm of nitrogen predominately in the form of $p_1$ center or donor nitrogen. The concentration of donor nitrogen was determined by IR spectroscopy. The main absorption peak is located at 1130 cm⁻¹, and is associated with the resonance oscillation of the N-C bond. The IR absorption spectra in our samples were recorded on an "Infralum FT-801" spectrophotometer (see Figure 5A). The absorption coefficient $\mu_C$ due to donor nitrogen was determined with respect to the intrinsic lattice absorption of the diamond. The concentrations of donor nitrogen were calculated using the formula [15,39]:

$$N_{p1}[\text{ppm}] = (25 \pm 2)\mu_C \tag{8}$$

Other important defects are NV⁻ (see below), N⁺, which could be found from 1332 cm⁻¹ absorption line using following formula [15]:

$$N_{N+}[\text{ppm}] = (5.5 \pm 1)\mu_{1332} \tag{9}$$

And interstitial nitrogen, concentration of which could be found as:

$$N_{in}[\text{ppm}] = (3.0 \pm 0.6)\mu_{1450} \tag{10}$$

Sum of all these color center gives full concentration of nitrogen in the sample. This color centers, nevertheless are just most common, but not all nitrogen-related centers. Therefore overall error on nitrogen concentration was estimated by measuring same sample before and after annealing at 800 °C, which given discrepancy up to 25% and been taken as our error bar.

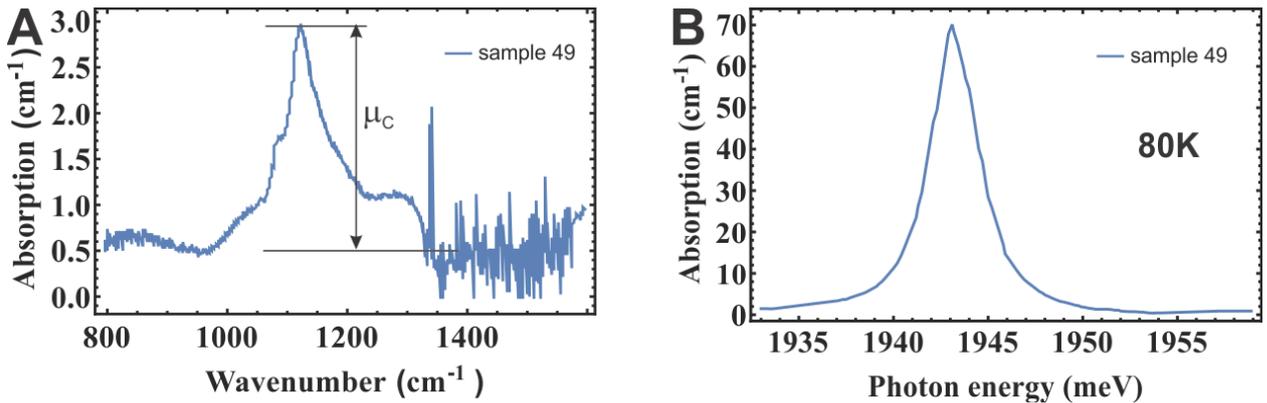

*Figure 5 A – IR spectra of the diamond plate 49, B – low temperature spectra of the sample 49*

*Processing of diamond plates*

In the next stage, the crystals were irradiated on a linear accelerator by electrons with an energy of 3 MeV and doses from $10^{17}$ to $10^{18}$ cm⁻². This led to the generation of isolated vacancies in the crystal structure of the diamond in neutral and negative charge states, with a total concentration, depending on the irradiation dose, from 5 to 9 ppm [15]. The subsequent annealing of diamonds at $800°C$, due to the mobility of vacancies, leads to the formation of NV defects.

If necessary, additional annealing is performed, the most useful and effective of which occurs at $1400°C$ for 24 hours. Such annealing parameters are the most optimal: on the one hand, the thermo-diffusion of atoms of the donor nitrogen with the formation of an A-defect: pairs of nitrogen atoms at neighboring lattice sites, is not yet activated; and on the other hand, at such temperatures, Ni and $N^+$ defects are already burned out and the lattice retains only the atoms of donor nitrogen and NV defects.

After all these procedures, the {111} orientation plates are cut from the single crystals and uniformly colored regions are selected on the cut plates, with regions with a uniform concentration of NV defects belonging to one crystal growth sector. Studied diamond plates were then made from them.

*Measurement of the concentration of NV centers in plates*

The concentration of donor nitrogen in the plates was measured from the intensity of the infrared absorption peaks before irradiation with electrons. After irradiation and annealing, the spectroscopy was carried out again. This also made it possible to estimate the concentration of $p_1$ defects in the plates. The concentration of NV centers in the first two plates was calculated by the formula [39]:

$$n_{NV[ppm]} = 0.0412 I_{637} \qquad (11)$$

$$I_{637\left[meV*cm^{-1}\right]} = \int_0^\infty \mu(E)dE, \qquad (12)$$

where $n_{NV[ppm]}$ is concentration of NV centers in ppm and $I_{637\left[meV*cm^{-1}\right]}$ – integral intensity of the zero-phonon line at 80 K (in the absorption spectra, see Figure 5B). Such a method is applicable only to sufficiently high concentrations of defects in a crystal. Therefore, the second group of plates with low concentrations was measured by another method and normalized to the measured concentration of the first plates. Using confocal microscope saturation curves of the NV center ensembles were measured (see Figure 3A). The concentration of NV⁻ centers was estimated via measuring the slope of the saturation curve and comparing it to that of the calibrated plates using the following formula:

$$n_{NV} = \frac{n_{NV\_norm} \alpha}{\alpha_{norm}} \qquad (13)$$

where $\alpha$ is the angle of slope of the saturation curve that was estimated on the measured plate; $\alpha_{norm}$ is the angle of the slope of the saturation curve that was estimated on the plate with a known concentration; $n_{NV\_norm}$ is the concentration of NV centers in the plate with a known concentration; and $n_{NV}$ is the concentration of NV centers in the measured plate. This method is valid as long as absorption of green laser by other defects in diamond on the length of the focal depth is not large, which was verified separately.

*Estimation of the collection efficiency with a parabolic lens concentrator*

To estimate the coupling efficiency in the experiment with a parabolic lens concentrator, a ray trace simulation was performed. The open source library LightPyCL, available from GitHub, was used for this purpose. The NV center was simulated as a point light source placed in the middle of a diamond of a size $1 \times 1 \times 0.3$ mm. In total, $2 \cdot 10^4$ rays were simulated for a single NV center. The ratio of the number of rays reaching the detector surface and the total emitted rays was used to find collection efficiency. This number needs to be corrected on number of factors: first, the exact position of the NV center will affect the result. To take this into account, 10 random points were selected inside the diamond plate and the same procedure was repeated for all the emitters, thus giving $\eta = 1305/20000 \approx 6.5\%$. This number needed to be further corrected on filters transmission, which in our setup for NV emission was about 95%. Finally, the transmission of the system was estimated as 6.2%

We note that this value is order of magnitude less than it was previously reported for CPC. Indeed, 65% could be obtained directly on the exit point of the CPC, while in our setup we have an air spacer of roughly 2.5 cm between CPC exit side and photodetector plate, which reduced the collection efficiency dramatically due to the large divergence of exit fluorescence (total angle 90 degrees).

*Measurement of coherent properties of NV ensembles in diamond*

The green laser (Coherent compass 315M 100 mW) was used as the pump laser. The laser beam was focused on an objective Olympus Plan 10X (Figure 1A) with a 1 cm long working distance, allowing positioning of the sample on the microwave antenna. For the microwave antenna, we used a resonant Helmholtz rings antenna, which provides a uniform but moderate magnetic field. The diameter of the rings is about 7 mm, and the antenna has resonance at 2920 MHz with FWHM of about 100 MHz. In addition, a constant magnetic field was provided to split the energy levels (Figure 1B) of the NV centers and pick only one of 4 possible orientation NV center. This was done by using a permanent magnet aligned with respect to the diamond's $\langle 111 \rangle$ crystallographic axis and was used to create a field along $\langle 111 \rangle$ axis. The NV centers' fluorescence was collected with a homemade confocal microscope and sent on a photodetector PerkinElmer APD (SPCM-AQRH-14-FC). We attenuated the beam with normal density filters to obtain 2-8 million photon counts on APD. National instruments PCIe-7851 FPGA card was used to produce control pulses for optical and microwave pumping. APD counting and gating is implemented internally in the card. Time resolution for all experiments was limited to 12.5 ns. Measurements of coherence times were organized in the following way. The sample was placed in a constant magnetic and alternating microwave magnetic field mostly applied perpendicular to $\langle 111 \rangle$ axis or Z axis of the microscope. A green laser was used to bring the system to the ground state.

Information about fluorescence was processed by the computer using National Instruments Labview software.

The optical Rabi oscillations were measured by applying a microwave pulse at the frequency of the transition from $m_s = 0$ to $m_s = 1$ and to the variation duration of the microwave pulse $\tau$ (see Figure 1C). The decay of the coherence oscillations was fitted with:

$$f(\tau) = a\cos(2\pi\omega_{Rabi}\tau + \varphi)e^{-\frac{\tau}{T^*_{2Rabi}}}. \tag{14}$$

Here, the fitting parameters were $a, \omega_{Rabi}, T^*_{2Rabi}, \varphi$. Where $T^*_{2Rabi}$ was taken as Rabi coherence time.

The Ramsey dephasing time $T^*_{2Ramsey}$ was estimated by the Ramsey sequence (see Figure 1C). The varying parameter was the delay $\tau$ between two $\pi/2$ pulses. The coherence decay curve was fitted by:

$$f(\tau) = e^{-\left(\frac{\tau}{T^*_{2Ramsey}}\right)^2} \sum_{i=1}^{3} a_i \cos(2\pi\Delta_i\tau + \varphi_i), \tag{15}$$

with fitting parameters $a, \Delta, T^*_{2Ramsey}, \varphi$ in which $T^*_{2Ramsey}$ was taken as Ramsey coherence time.

The coherence time $T_2$ was estimated by the Hahn echo sequence [40]; see Figure 1C. The varying parameter was the delay $\tau$ between $\pi/2$ and $\pi$. The decay of echo signal was fitted by:

$$f(\tau) = ae^{-\left(\frac{2\tau}{T_2}\right)^\alpha}, \tag{16}$$

where $a, \alpha, T_2$ are fitting parameters and $T_2$ is was taken as the Echo coherence time.

In case of low concentration of nitrogen one could observe revivals in coherence decay at large enough magnetic fields. In our experiments the maximum magnetic field was around 86 G the revivals were only seen for low concentration of nitrogen in the plates. In this case, following [41] we used following fitting formula:

$$f(\tau) = ae^{-\left(\frac{2\tau}{T_2}\right)^\alpha}(1+\cos[\omega t])^\beta, \tag{17}$$

where $\beta$ is empirical fitting parameter.

*Strain measurements of diamond plates*

To estimate the strain parameters of the diamond plates, we performed ODMR measurements of a small ensemble of NV centers in the absence of a static magnetic field. The laboratory magnetic field was compensated by applying a magnetic field equal and opposite in direction to the laboratory magnetic. Using the formalism given in [42], the positions of individual hyperfine lines could be determined analytically given the value of applied strain and magnetic field. The system is more susceptible to transverse strain than to longitudinal. In that approximation the ODMR spectrum of single NV center could be expressed as:

$$S_{NV}(f, f_0, \Gamma) = \frac{1}{2} L(f, D_{ZFS} \pm \mathrm{E}_\perp, \Gamma) + L(f, D_{ZFS} \pm \sqrt{A_\parallel^2 + \mathrm{E}_\perp^2}, \Gamma), \qquad (18)$$

where $L(x, x_0, \gamma)$ is a Lorentzian function of the form $L(x, x_0, \gamma) = \gamma^2 / \left[(x - x_0)^2 + \gamma^2\right]$; $\mathrm{E}_\perp$ is the effective energy of interaction with the strain fields (see, for example, [43] for details); $f$ is swept MW frequency; $f_0$ is the resonance frequency; $\Gamma$ is the linewidth of the ODMR spectrum; $D_{ZFS}$ is 2.87 MHz zero field splitting; $A_\parallel$ is hyperfine constant; and $\mathrm{E}_\perp$ is the energy of interaction with transverse strain fields.

To account for inhomogeneous broadening, we assumed that strain is distributed with a Gaussian distribution:

$$\mathrm{E}_\perp(x) = \frac{1}{2\pi\sigma} Exp[-\frac{(x - \Pi)^2}{2\sigma^2}]. \qquad (19)$$

where $\Pi$ is an average strain field, $x$ is the local strain field, and $\sigma$ is the width of the distribution. The ODMR spectrum (18) was then convolved with (19). Obtained expression for strain was numerically evaluated in Wolfram Mathematica software. Parameters of the expression were adjusted manually to best fit inside the feature of the ODMR spectrum as most sensitive to strain in the lattice. This approach results in good agreement with experimental data. At some plates with low strain and ODMR linewidth, however, the outer wings of the ODMR spectrum deviate to form the fit. This is most likely due to the presence in our plates of a natural abundance of $^{13}$C isotopes, which give rise to additional broadening.

# References

[1]   T. Schröder, F. Gädeke, M. J. Banholzer, and O. Benson, New J. Phys. **13**, 055017 (2011).

[2]   J. Joo and E. Ginossar, Sci. Rep. **6**, 26338 (2016).

[3]   H. Bernien, B. Hensen, W. Pfaff, G. Koolstra, M. S. Blok, L. Robledo, T. H. Taminiau, M. Markham, D. J. Twitchen, L. Childress, and R. Hanson, Nature **497**, 86 (2013).

[4]   G. Kucsko, P. C. Maurer, N. Y. Yao, M. Kubo, H. J. Noh, P. K. Lo, H. Park, and M. D. Lukin, Nature **500**, 54 (2013).

[5]   J. M. Taylor, P. Cappellaro, L. Childress, L. Jiang, D. Budker, P. R. Hemmer, A. Yacoby, R. Walsworth, and M. D. Lukin, Nat. Phys. **4**, 810 (2008).

[6]   D. Le Sage, K. Arai, D. R. Glenn, S. J. DeVience, L. M. Pham, L. Rahn-Lee, M. D. Lukin, A. Yacoby, A. Komeili, and R. L. Walsworth, Nature **496**, 486 (2013).

[7]   A. Ajoy, U. Bissbort, M. D. Lukin, R. L. Walsworth, and P. Cappellaro, Phys. Rev. X **5**, 011001

(2015).

[8]  J. Wang, F. Feng, J. Zhang, J. Chen, Z. Zheng, L. Guo, W. Zhang, X. Song, G. Guo, L. Fan, C. Zou, L. Lou, W. Zhu, and G. Wang, Phys. Rev. B - Condens. Matter Mater. Phys. **91**, 1 (2015).

[9]  P. Ovartchaiyapong, K. W. Lee, B. A. Myers, and A. C. B. Jayich, Nat. Commun. **5**, 4429 (2014).

[10] D. Farfurnik, N. Alfasi, S. Masis, Y. Kauffmann, E. Farchi, Y. Romach, Y. Hovav, E. Buks, and N. Bar-Gill, Appl. Phys. Lett. **111**, 123101 (2017).

[11] E. Bauch, C. A. Hart, J. M. Schloss, M. J. Turner, J. F. Barry, P. Kehayias, S. Singh, and R. L. Walsworth, Phys. Rev. X (2018).

[12] G. de Lange, T. van der Sar, M. Blok, Z.-H. Wang, V. Dobrovitski, and R. Hanson, Sci. Rep. **2**, 382 (2012).

[13] F. Fávaro de Oliveira, D. Antonov, Y. Wang, P. Neumann, S. A. Momenzadeh, T. Häußermann, A. Pasquarelli, A. Denisenko, and J. Wrachtrup, Nat. Commun. **8**, 15409 (2017).

[14] K. Fang, V. M. Acosta, C. Santori, Z. Huang, K. M. Itoh, H. Watanabe, S. Shikata, and R. G. Beausoleil, Phys. Rev. Lett. **110**, 130802 (2013).

[15] I. A. Dobrinets, V. G. Vins, and A. M. Zaitsev, *HPHT-Treated Diamonds*, Springer s (Springer, 2008).

[16] N. Mizuochi, P. Neumann, F. Rempp, J. Beck, V. Jacques, P. Siyushev, K. Nakamura, D. J. Twitchen, H. Watanabe, S. Yamasaki, F. Jelezko, and J. Wrachtrup, Phys. Rev. B **80**, 041201 (2009).

[17] V. M. Acosta, E. Bauch, M. P. Ledbetter, C. Santori, K.-M. C. Fu, P. E. Barclay, R. G. Beausoleil, H. Linget, J. F. Roch, F. Treussart, S. Chemerisov, W. Gawlik, and D. Budker, Phys. Rev. B **80**, 115202 (2009).

[18] G. Kucsko, S. Choi, J. Choi, P. C. Maurer, H. Sumiya, S. Onoda, J. Isoya, F. Jelezko, E. Demler, N. Y. Yao, and M. D. Lukin, (2016).

[19] S. Hong, M. S. Grinolds, L. M. Pham, D. Le Sage, L. Luan, R. L. Walsworth, and A. Yacoby, **38**, 155 (2013).

[20] A. Stark, N. Aharon, T. Unden, D. Louzon, A. Huck, A. Retzker, U. L. Andersen, and F. Jelezko, Nat. Commun. **8**, 1105 (2017).

[21] D. Farfurnik, A. Jarmola, L. M. Pham, Z. H. Wang, V. V Dobrovitski, R. L. Walsworth, D. Budker, and N. Bar-Gill, Phys. Rev. B **92**, (2015).

[22] K. N. Shinde, S. J. Dhoble, H. C. Swart, and K. Park, *Phosphate Phosphors for Solid-State Lighting*, Springer s (Springer, 2012).

[23] J. Schwartz, S. Aloni, D. F. Ogletree, M. Tomut, M. Bender, D. Severin, C. Trautmann, I. W. Rangelow, and T. Schenkel, J. Appl. Phys. **116**, 214107 (2014).

[24] C. A. McLellan, B. A. Myers, S. Kraemer, K. Ohno, D. D. Awschalom, and A. C. Bleszynski

Jayich, Nano Lett. **16**, 2450 (2016).

[25] A. J. Neves and Maria Helena Nazaré, editors , *Properties, Growth and Applications of Diamond* (INSPEC, The Institution of Electrical Engineers , London, United Kingdom, 2001).

[26] A. O. Levchenko, V. V. Vasil'Ev, S. A. Zibrov, A. S. Zibrov, A. V. Sivak, and I. V. Fedotov, Appl. Phys. Lett. **106**, (2015).

[27] A. T. Collins, L. Allers, C. J. H. Wort, and G. A. Scarsbrook, Diam. Relat. Mater. **3**, 932 (1994).

[28] N. Tobias, K. Buczak, A. Angerer, S. Putz, G. Steinhauser, H. Peterlik, J. Majer, and M. Trupke, 1 (2013).

[29] V. V. Vorobyov, V. V. Soshenko, S. V. Bolshedvorskii, J. Javadzade, N. Lebedev, A. N. Smolyaninov, V. N. Sorokin, and A. V. Akimov, Eur. Phys. J. D **70**, 269 (2016).

[30] V. V Soshenko, O. R. Rubinas, V. V Vorobyov, S. V Bolshedvorskii, P. V Kapitanova, V. N. Sorokin, and A. V Akimov, Bull. Lebedev Phys. Inst. **45**, 237 (2018).

[31] M. W. Doherty, F. Dolde, H. Fedder, F. Jelezko, J. Wrachtrup, N. B. Manson, and L. C. L. Hollenberg, **205203**, 1 (2012).

[32] H. Clevenson, M. E. Trusheim, T. Schroder, C. Teale, D. Braje, and D. Englund, Nat. Phys. **11**, 393 (2015).

[33] D. R. Glenn, D. B. Bucher, J. Lee, M. D. Lukin, H. Park, and R. L. Walsworth, Nature **555**, 351 (2018).

[34] K. D. Jahnke, B. Naydenov, T. Teraji, S. Koizumi, T. Umeda, J. Isoya, and F. Jelezko, Appl. Phys. Lett. **101**, 012405 (2012).

[35] V. Stepanov and S. Takahashi, Phys. Rev. B **94**, 1 (2016).

[36] I. P. Radko, M. Boll, N. M. Israelsen, N. Raatz, J. Meijer, F. Jelezko, U. L. Andersen, and A. Huck, Opt. Express **24**, 27715 (2016).

[37] Y. Ruan, D. A. Simpson, J. Jeske, H. Ebendorff-Heidepriem, D. W. M. Lau, H. Ji, B. C. Johnson, T. Ohshima, S. V. Afshar, L. Hollenberg, A. D. Greentree, T. M. Monro, and B. C. Gibson, Sci. Rep. **8**, 1 (2018).

[38] T. Wolf, P. Neumann, K. Nakamura, H. Sumiya, T. Ohshima, J. Isoya, and J. Wrachtrup, Phys. Rev. X **5**, 041001 (2015).

[39] S. C. Lawson, D. Fisher, D. C. Hunt, and M. E. Newton, J. Phys. Condens. Matter **10**, 6171 (1998).

[40] E. L. Hahn, Phys. Rev. **80**, 580 (1950).

[41] P. L. Stanwix, L. M. Pham, J. R. Maze, D. Le Sage, T. K. Yeung, P. Cappellaro, P. R. Hemmer, A. Yacoby, M. D. Lukin, and R. L. Walsworth, Phys. Rev. B **82**, 201201 (2010).

[42] M. W. Doherty, F. Dolde, H. Fedder, F. Jelezko, J. Wrachtrup, N. B. Manson, and L. C. L. Hollenberg, Phys. Rev. B **85**, 205203 (2012).

[43] M. E. Trusheim and D. Englund, New J. Phys. **18**, 123023 (2016).